# `piecewiseSEM`: Piecewise structural equation modeling in R for ecology, evolution, and systematics

Jonathan S. Lefcheck[1†]

[1]Department of Biological Sciences, Virginia Institute of Marine Science, The College of William & Mary, Gloucester Point, VA 23062-1346, USA

[†]Corresponding author: jslefche@vims.edu. Mailing address: PO Box 1346, Rt 1208 Greate Rd, Gloucester Point, VA 23062, USA

**Abstract**

1. Ecologists and evolutionary biologists are relying on an increasingly sophisticated set of statistical tools to describe complex natural systems. One such tool that has gained increasing traction in the life sciences is structural equation modeling (SEM), a variant of path analysis that resolves complex multivariate relationships among a suite of interrelated variables.

2. SEM has historically relied on covariances among variables, rather than the values of the data points themselves. While this approach permits a wide variety of model forms, it limits the incorporation of detailed specifications.

3. Here, I present a fully-documented, open-source R package *piecewiseSEM* that builds on the base R syntax for all current generalized linear, least-square, and mixed effects models. I also provide two worked examples: one involving a hierarchical dataset with non-normally distributed variables, and a second involving phylogenetically-independent contrasts.

4. My goal is to provide a user-friendly and tractable implementation of SEM that also reflects the ecological and methodological processes generating data.

"No aphorism is more frequently repeated in connection with field trials, than that we must ask Nature few questions, or, ideally, one question, at a time. The writer is convinced that this view is wholly mistaken. Nature, he suggests, will best respond to a logical and carefully thought out questionnaire; indeed, if we ask her a single question, she will often refuse to answer until some other topic has been discussed." –Sir Ronald Fisher (1926)

**Introduction**

Nature is complicated. This may seem like an obvious statement, but the desire to understand the complexity of nature is arguably the single driving force behind all of science. Yet, for most of its history, ecology and evolutionary biology has been largely concerned with stripping away all extraneous influences and closely examining the impact of one or few factors on a single response. This practice was and sometimes still is a consequence of limited computational power, and the necessity of simplification in rigorous experimentation. However, with the advent of modern computing and the tractability of large-scale observational surveys, there is an increasing recognition that multifaceted datasets representing complex natural systems require an equally sophisticated toolbox to help uncover the causal relationships among suites of interconnected variables. Structural equation models, or SEMs, provide one such tool.

Structural equation models are probabilistic models that unite multiple predictor and response variables in a single causal network (Grace 2006). SEMs are often represented using path diagrams, where arrows indicate directional relationships between observed variables (Fig. 1, 2). These relationships can be captured in a series of structure equations that correspond to the pathways in the model. Two primary characteristics of SEMs separate them from more traditional modeling approaches:

(1) Paths represent <u>hypothesized causal relationships</u>. This is a departure from the mantra 'correlation does not imply causation.' In fact, correlation does imply causation, but the direction of causality is unresolved, since we cannot know whether A causes B, B causes A, or both A and B are a consequence of some third, unmeasured variable (Shipley 2000b). By using pre-existing knowledge of the system gained through observation and/or experimentation, we can make an informed hypothesis about the causal structure of A, B, and other variables that are thought to mediate their relationship. SEM allows for the direct test of this supposed causal structure. In this way, SEM is a departure from traditional linear modeling by being said to test the hypothesis that A *causes* B. While an in-depth discussion of causality is beyond the scope of this paper, further discussion of causality and how it relates to SEM can be found in: Pearl (2012) and Bollen & Pearl (2013).

(2) Variables can appear as <u>both</u> predictors and responses. By allowing one variable to serve as a response in one path and as a predictor in another, SEM is useful for testing and quantifying indirect or cascading effects that would otherwise go unnoticed in any single model (e.g., Grace et al., 2007).

SEM thus requires a sea change in how ecological and evolutionary questions are structured and tested, with a shift towards the simultaneous evaluation of multiple causal hypotheses within a single network.

Historically, SEMs have been estimated using a maximum-likelihood approach to select parameter values that best reproduce the entirety of the observed variance-covariance matrix. The goodness-of-fit of the SEM can then be evaluated using a $\chi^2$ test comparing the estimated to the observed covariance matrix (Grace 2006). This approach, however, assumes that all observations are independent, and all variables follow a (multivariate) normal distribution (Grace 2006). It also restricts the minimum number of observations necessary to fit the SEM, since there needs to be sufficient degrees of freedom to estimate the whole variance-covariance matrix (the '*t* rule', (Grace 2006).

These restrictions led to the parallel development of directed acyclic, or piecewise, SEMs based on applications from graph theory. In piecewise SEM, the path diagram is translated to a set of linear equations, which are then evaluated individually. The switch from *global estimation*, where equations are solved simultaneously, to *local estim*ation, where each equation is solved separately, allows for the fitting of a wide range of distributions and sampling designs (Shipley 2000a, 2009). It also, in theory, permits the fitting of smaller datasets, since there only need be enough degrees of freedom to fit any given component model (Shipley 2000a) (but see Discussion: *Limitations*). Finally, it can incorporate distances obtained from taxonomy or phylogeny to address the potentially confounding effects of shared evolutionary history (von Hardenberg & Gonzalez-Voyer 2013).

Since piecewise SEM produces no valid global covariance matrix, alternative goodness-of-fit tests are necessary. The typical approach uses Shipley's test of *directed separation*. This procedure tests the assumption that all variables are conditionally independent. In simplest terms, conditional independence implies that there are no missing relationships among unconnected variables (Shipley 2000a). The first step in the test of direct separation is to derive the minimum set of conditional independence claims associated with the hypothesized path diagram, known as the *basis set*. The basis set can be translated into a set of linear equations, each of which can be solved like any other linear model. The significance of any given independence claim, i.e., its *P*-value, can be estimated and extracted. The test of directed separation is conducted by combining all *P*-values across the basis set in a test statistic, Fisher's *C*, using the following equation:

$$C = -2 \sum_{i=1}^{k} \ln(P_i) \quad (1)$$

where $P_i$ is the *i*th independence claim in a basis set consisting of *k* claims. *C* can then be compared to a $\chi^2$-distribution with 2*k* degrees of freedom. The hypothesized relationships are considered to be consistent with the data when there is weak support for the sum of the conditional independence claims, i.e., where the collection of such relationships represented by *C* could have easily occurred by chance, in which case *P* for the $\chi^2$ test is greater than the chosen significance threshold (typically α = 0.05). Several approachable examples of the derivation of basis sets can be found in Shipley (2000a, 2009).

Shipley (2013) showed that the Fisher's *C* statistic can be used to obtain a value of Akaike's Information Criterion (AIC) using the following equation:

$$AIC = C + 2K \qquad (2)$$

Where *C* is from Equation (1), and *K* is the likelihood degrees of freedom (not to be confused with *k*, the number of independence claims in the basis set). Equation (2) can also be extended to small sample sizes ($AIC_c$, typically when the number of parameters exceeds the total sample size *n*/40) using an additional correction: $2K(\frac{n}{n-K-1})$. Because this estimator is not derived from maximum-likelihood, it is sometimes referred to as the *C* statistic Information Criterion (CIC, *sensu* Cardon *et al.* 2011).

The implementation of piecewise SEM is limited by the correct specification and evaluation of the basis set, which can be prohibitive to obtain by hand, especially for very complex models. To that end, I provide a fully-documented and open-source R package (R Development Core Team 2015) called *piecewiseSEM*[1] to aid in the calculation of piecewise structural equation models by: constructing the basis set, conducting goodness-of-fit tests for both the full and component models, calculating AIC scores, returning (scaled) parameter estimates, plotting partial correlations, and generating predictions. SEMs are specified using a list of structured equations, which can be built using most common linear modeling approaches in R and thus can accommodate non-normal distributions, hierarchical structures, and different estimation procedures. I also extend the tests of directed separation to include interactions. In this paper, I present two worked examples: the first incorporating generalized hierarchical mixed effects models, and the second incorporating phylogenetically-independent contrasts. The data and R code to reproduce all analyses are given in the supplements and at: https://github.com/jslefche/piecewiseSEM_examples.

---

[1] https://github.com/jslefche/piecewiseSEM

**Example 1: Storm frequency and kelp forest food webs**

In this first example, I use data from Byrnes *et al.* (2011), who examined the role of storm events on the diversity and food web structure of kelp forests in California, USA. They combined biological surveys of kelp forests over 35 different sites and 8 years, potential food-web linkages derived from the literature, data on wave height and period from physical monitoring stations, and kelp canopy cover from satellite imagery. They summarized these variables in a single causal network derived using *a priori* knowledge of the system and results from experimental manipulation (Fig. 1). They then evaluated this model using traditional variance-covariance SEM. They hypothesized that the wave disturbance generated by winter storms would be contingent on the amount of existing kelp, which interactively affect the spring canopy cover. Spring canopy cover would in turn inform summer canopy cover, which would also be subject to physical forcing. The amount of canopy cover, spring or summer, would provide structural habitat for various species, such as algae, sessile invertebrates, and their consumers. Total species richness would in turn determine the number of potential trophic links in the observed food web (linkage density, or the mean number of feeding links per observed species).

The results of their original analysis are reproduced in Figure 1a using the *lavaan* package (Rosseel 2012). The model fit the data adequately based on output from a $\chi^2$ goodness-of-fit test ($\chi^2_5 = 8.784, P = 0.118$). Byrnes *et al.* (2011) saw that spring canopy cover was strongly influenced by the interaction between wave disturbance and previous kelp cover: as the previous year's cover increased, the effect of wave disturbance on the current spring's canopy cover became more negative. Spring canopy cover had both a direct negative effect on species richness, and an indirect positive effect mediated through summer kelp cover. Species richness in turn enhanced food web complexity. However, they noted that the direct negative effect of spring canopy cover on species richness had a larger magnitude (-0.23) vs. the indirect effect, which can be obtained by multiplying the path coefficients (0.38 * 0.29 = 0.11). Thus, they concluded that the removal of spring canopy by winter storms actually increased species richness (by reducing the stronger direct negative effect), ultimately increasing food web complexity in the short term. However, given the effect of losing kelp, total species richness should decline if reefs experienced multiple years of wave disturbance in a row.

Their analysis, however, treated each observation as independent. In reality, sites that are proximate are likely to share similar characteristics, and within a site, observations closer in time are likely to be more similar than those that are farther apart. To address both of these concerns, I re-fit their original model using piecewise SEM. In the first re-analysis, I addressed the non-independence of sampling sites by fitting each response to a general linear mixed effects models using the function `lme` from the *nlme* package (Pinheiro *et al.* 2013). I chose to log-transform the variables as in Byrnes *et al.* (2011) instead of fitting integer responses in order to facilitate direct comparisons to the original analysis. For each component model, I fit a random effect of *Site* and allowed only its intercept to vary. I then added the component models to a list and passed the list to the function `sem.fit`, which returns the tests of directed separation, Fisher's *C* statistic and AIC values for the SEM. I then recovered the standardized regression coefficients (scaled by mean and variance, as in Byrnes *et al.*) using the `sem.coefs` function.

The piecewise SEM based on mixed models reproduced the data well based on comparison of the Fisher's *C* statistic to a $\chi^2$ distribution ($C_{10} = 15.64, \mathrm{P} = 0.11$). The results from this re-analysis are given in Figure 1b. In general, the models explained a larger proportion of variance on average than the traditional SEM, based on $R^2$ values derived from the variance of both fixed and random effects (Nakagawa & Schielzeth 2012) obtained using `sem.model.fits`. There are several major differences between the models in Figure 1a and 1b. First, wave disturbance has a significant negative effect on spring canopy cover in the absence of kelp cover (-0.22), where it was previously non-significant. The magnitudes of the main effect of the previous year's kelp canopy cover and the interaction between this variable and wave disturbance are both reduced by about two-thirds, although they retain the same signs. Most consequential for the original interpretation is that the negative relationship between spring canopy cover and species richness is now non-significant. By nesting observations based on their hierarchical structure, variation that was formerly assumed to be generated by canopy cover was reallocated to random (spatial) variation. Thus, based on the output from the piecewise SEM, I must infer that wave disturbance both directly and indirectly reduces spring canopy cover, which now indirectly *reduces* food web complexity as a consequence of cascading positive relationships between spring and summer canopy cover, summer canopy cover and species richness, and finally species richness and linkage density.

In the second re-analysis, I addressed both the non-independence of sites as well as any potential temporal autocorrelation by retaining the same random structure as above, and additionally modeling the correlation among sampling years using a continuous autoregressive 1 autocorrelation structure from the `CAR1` function from the *nlme* package (Pinheiro *et al.* 2013). This analysis also reproduced the data well based on comparison of the Fisher's *C* statistic to a $\chi^2$ distribution ($C_8 = 7.84, \mathrm{P} = 0.45$). The results from this re-analysis are given in Figure 1c. There is slightly greater amount of variance explained for each component model versus the piecewise SEM without the autocorrelation structure. There are, however, fewer notable differences between the two piecewise models. The path between spring canopy cover and species richness is still non-significant. There is now a significant positive path between summer canopy cover and wave disturbance, and the formerly significant path between the previous year's canopy cover and species richness is now non-significant. However, comparison of the two piecewise SEMs using AIC reveals that the model additionally incorporating the CAR1 autocorrelation structure is considerably less likely model than the one with only the hierarchical random structure ($AIC_c$ = 97.69 vs. 81.44).

In sum, this re-analysis has revealed that modeling the hierarchical structure of the data leads to a different interpretation of the original data: wave disturbance *decreases* food web complexity, principally by removing habitat. This interpretation, however, supports the overall conclusions of Byrnes *et al.* (2011) that repeated storm events (i.e., wave disturbance) should decrease food web complexity, although I show this effect is mediated through the removal of habitat upon the first occurrence of disturbance and not necessarily a decrease in species richness after repeated disturbance events, as suggested by Byrnes *et al.* (2011). Additionally, AIC model comparisons revealed that modeling potential temporal autocorrelation does not add to our ability to understand this system of interactions.

Further exploration of the models from Byrnes *et al.* (2011) decomposing total species richness into trophic components revealed that canopy cover significantly reduced algal but not sessile invertebrate or mobile consumer species richness, as in their original analysis (see supplementary code). Modeling the random effect of *Site* likely absorbed some of the variation in algae-rich vs. algae-poor sites, making it more difficult to see the algae richness contribution to total species richness in the simpler piecewise model (Fig. 1b, c). This additional analysis confirms that the deeper exploration by Byrnes *et al.* (2011) was warranted to reconcile the statistical output with the biology of the system. While ultimately we arrive at the same conclusion, this example demonstrates the importance of honoring the way in which the data were collected, within the limits of the tools available at the time.

**Example 2: Eusociality and ecological success in sponge-dwelling shrimp**

In this second example, I use population and ecological data from a genus of sponge-dwelling shrimps, *Synalpheus*, to explore the drivers of ecological success. Species in this genus exhibit a range of social structures, from pair-forming to truly eusocial, with a single reproducing female per colony. It has been hypothesized that complex social structures like those exhibited by certain *Synalpheus* species are ecologically advantageous in fostering greater competitive ability and/or resource acquisition. To answer this question, Duffy & Macdonald (2010) collated data on female body mass, number of host species used (host range), and proportional regional abundance for 20 species of *Synalpheus* in Belize. They additionally calculated an index of eusociality for each species. They hypothesized that more eusocial species (i.e., larger colonies with a single breeding female) would occupy a wider range of hosts and be more successful in defending those hosts (i.e., achieve higher relative abundance in the study area). They additionally hypothesized that this effect might be confounded by body size, since most eusocial species are small-bodied.

As a first pass, I fit a traditional SEM using the `sem` function from the *lavaan* package (Rosseel 2012), assuming independence among all 20 data points (species). The model reproduced the data well ($\chi^2_1 = 0.653, P = 0.419$), and the results are given in Figure 2a. There are two significant paths of interest: a strong positive effect of eusociality on host range accounting for body mass (standardized regression coefficient = 0.58), and a positive effect of host range and relative abundance (0.47). There was not, however, a significant direct relationship between eusociality and abundance. Thus, it appears that the success of eusocial species is largely a consequence of their ability to occupy a wide range of hosts. As a consequence of this generalist habitat use, they then also make up a larger percentage of total regional abundance, but the model does not support the hypothesis that eusociality confers a direct advantage in defending and holding onto a particular habitat resource.

Of course, Duffy & Macdonald (2010) correctly point out that the data points are not independent because some species are more related than others. To rectify this issue, I re-fit the SEM in Figure 1a but additionally fixed the model correlation matrix based on genetic distances derived from a phylogeny of *Synalpheus* in the region (Hultgren & Duffy 2012). I obtained the model correlations from the phylogenetic tree using the function `corBrownian` from the *ape* package (Paradis *et al.* 2004), and fit the component models using the function `gls` from the *nlme* package (Pinheiro *et al.* 2013). I stored the component models in a list and then evaluated the SEM using `sem.fit`. As before, the model reproduced the data well ($C_8 = 0.57, \text{P} = 0.751$), and the results are given in Figure 2b. The striking difference between the two SEMs in Figure 2 is that the phylogenetic piecewise SEM recovers a significant negative effect of body mass on host range (-0.32), supporting the expectation that body size has a confounding influence. Even in the presence of a body size effect, there is a significant positive effect of eusociality on host range (indeed it is substantially stronger, 0.80). As with the previous SEM (Fig. 1a), there was no direct effect of eusociality on proportional regional abundance. Again, this relationship was mediated through an increase in host range.

In their original paper, Duffy & Macdonald (2010) used multiple linear regression to explore relationships among these four variables. In their analysis, they showed that eusociality had a strong positive relationship with both relative abundance and host range size, after accounting for differences in body size and shared evolutionary history. Here, in a re-analysis of their data using SEM, I show the relationship between eusociality and relative abundance is not direct, but rather an indirect consequence of occupying a wider number of hosts, an insight that was not possible to infer from the multiple regressions. While tools like phylogenetic least-squares regression have been used for decades, the extension of these methods to SEM facilitates the testing of more complex, multivariate – and thus, realistic – hypotheses in evolutionary ecology.

**Discussion**

In this paper, I briefly introduce the concepts behind piecewise structural equation modeling (SEM), and apply piecewise SEM to two existing analyses. In both cases, acknowledging the non-independence of data points by incorporating random variation or phylogenetic contrasts yielded substantially different inferences than multiple regression or even traditional variance-covariance SEM. I also demonstrate how a new R package, *piecewiseSEM*, can be used to quickly and easily implement complex local estimation. Indeed, this package has already been used to explore the planetary drivers of ecosystem functioning in eelgrass beds (Duffy *et al.* 2015), disentangle the influence of functional diversity across trophic levels in experimental estuarine mesocosms (Lefcheck & Duffy 2015), and quantify the biotic and abiotic drivers of grassland multifunctionality (Jing *et al.* 2015).

*Broader Applications*

The piecewise SEM package contains a number of additional functions that may be of general interest. `sem.model.fits`, for example, generates $R^2$ and pseudo-$R^2$ and AIC values for component models based on methods in (Nakagawa & Schielzeth 2012; Johnson 2014). `predict.sem` is a wrapper for the generic `predict` function and additionally implements standard errors on predictions from models constructed using `lmer` based on the variance of the fixed effects only (from: http://glmm.wikidot.com/faq). `partial.resid` returns the partial correlation plot between two variables in a single model having accounted for the effects of covariates, and is an intuitive graphical way to emphasize the partial regression coefficients returned from `sem.coefs` or, more generally, `summary`. Exploration of partial correlations also allows for the identification of previously unrecognized non-linear relationships, which can then be incorporated into the model structure.

*Limitations*

While it has been suggested that piecewise SEM can be used to circumvent restrictions on sample size (Shipley 2000a), it is important to note that small sample sizes may still have severe consequences for the analysis. In particular, tests of directed separation may yield a 'good fitting model' only because the tests lacked sufficient power to uncover any actual trends. This outcome would be increasingly common as models increase in complexity. Ideally, investigators should devise the hypothesized model beforehand and use it to inform data collection, ensuring sufficient replication from the start. As a general rule, Grace *et al.* (2015) propose that the ratio of the total number of samples to the number of variables (*d*) should not fall below *d* = 5. It is also critical to examine the fits of the component models: if the overall SEM has an adequate fit but the component models have low explanatory power, then it is not acceptable (or useful) to draw inferences from the SEM. Finally, users may find themselves with the opposite problem, where large sample sizes cause everything to be significant at $\alpha = 0.05$. In this case, implementing a more stringent cutoff for statistical significance may alleviate the issue.

It is worth noting that *P*-values derived from the *lmerTest* package (Kuznetsova *et al.* 2013) are somewhat unstable at the time of writing, and can often lead to errors in the `sem.fit` function. Estimates from *nlme* appear to be more reliable, and I recommend users should construct their models using *nlme* when *lmerTest* produces an error.

While the piecewise SEM approach represents a considerable leap forward in addressing the assumptions of real-world data, its infancy relative to traditional SEM has led to some limitations. For instance, there is no real implementation of correlated errors, or relationships that are bidirectional and assumed to be caused by a shared underlying driver. *piecewiseSEM* implements a crude approximation of correlated errors by allowing the user to exclude them from the basis set (since there is no presumed direction of causality), and then running a simple test of significance on the bivariate correlation. Piecewise SEM also cannot disentangle cyclic relationships (e.g., A -> B -> C -> A), making it impossible to evaluate feedbacks (Shipley 2009). Similarly, the piecewise SEM method cannot evaluate reciprocal relationships in the same model (A -> B and B -> A, not to be confused with a bidirectional arrow indicating a correlated error). Finally, there is no formal integration of latent variables – those that are not directly measured, but inferred through a combination of observed variables (Grace 2006) – into piecewise SEM as of yet. It would be possible to derive predictions approximating a latent variable using exploratory factor analysis, or through the application of MCMC estimation. However, there has yet to be a thorough investigation and application of factor analysis to piecewise SEM. With luck, future developments will relax some of these limitations.

**Acknowledgments**

I am grateful to JEK Byrnes, JE Duffy and KM Hultgren for their generous sharing of the data used in this paper and comments on this manuscript. JE Duffy, JEK Byrnes, and J Grace provided comments on earlier drafts of the manuscript. Additionally, I thank A von Hardenberg for thoughtful discussion of phylogenetic SEM. Finally, I am indebted to the many folks, including those mentioned above, who have helped in developing and testing the *piecewiseSEM* package. Of particular additional note: N Deguines, TM Anderson, and T Mason. This paper is contribution no. #### of the Virginia Institute of Marine Science.

**Figure Captions:**

**Figure 1:** Structural equation model from Byrnes *et al.* (2011) exploring the effects of storm frequency (wave disturbance) on kelp forest community structure and food web complexity (linkage density). Boxes represent measured variables. Arrows represent unidirectional relationships among variables. Black arrows denote positive relationships, and red arrows negatives ones. Arrows for non-significant paths ($P \geq 0.05$) are semi-transparent. The thickness of the significant paths have been scaled based on the magnitude of the standardized regression coefficient, given in the associated box. $R^2$s for component models are given in the boxes of response variables (for panels **b** and **c**, this is reported as the conditional $R^2_c$ based on the variance of both the fixed and random effects). The variable "Reef habitat" has been omitted for clarity and the path coefficient is instead reported in the corresponding box of the response, as in Byrnes *et al.* (2011). **(a)** Original analysis using variance-covariance SEM. **(b)** The same model as in **a** fit using piecewise SEM and incorporating a nested random effect of *Transect* within *Site*. **(c)** The piecewise model from panel **b**, with an additional autocorrelation term for *Year*.

**Figure 2:** Structural equation model derived from hypotheses in Duffy & Macdonald (2010) exploring the relationships among eusociality, body size, host range size, and proportional regional abundance. Arrows represent unidirectional relationships among variables. Black arrows denote positive relationships, and red arrows negatives ones. Arrows for non-significant paths ($P \geq 0.05$) are semi-transparent. The thickness of the significant paths have been scaled based on the magnitude of the standardized regression coefficient, given in the associated box. $R^2$s for component models are given in the boxes of response variables; these values are omitted for panel **b** because there is no agreed coefficient of determination for `gls` models. **(a)** Analysis using variance-covariance SEM. **(b)** The same model as in **a** fit using piecewise SEM and incorporating a fixed correlation structure based on phylogenetic distances between species derived from a molecular phylogeny.

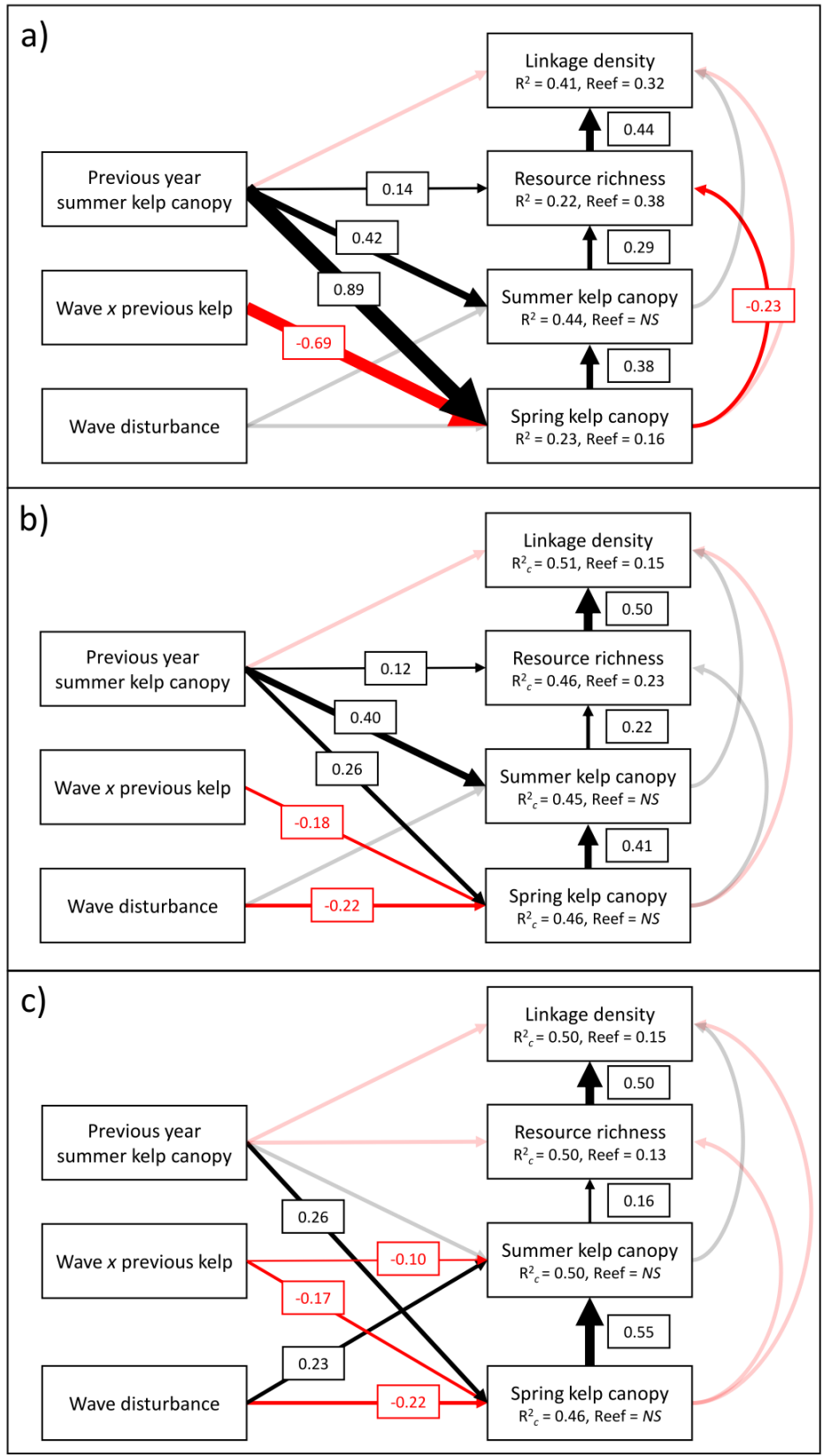

**Figure 1**

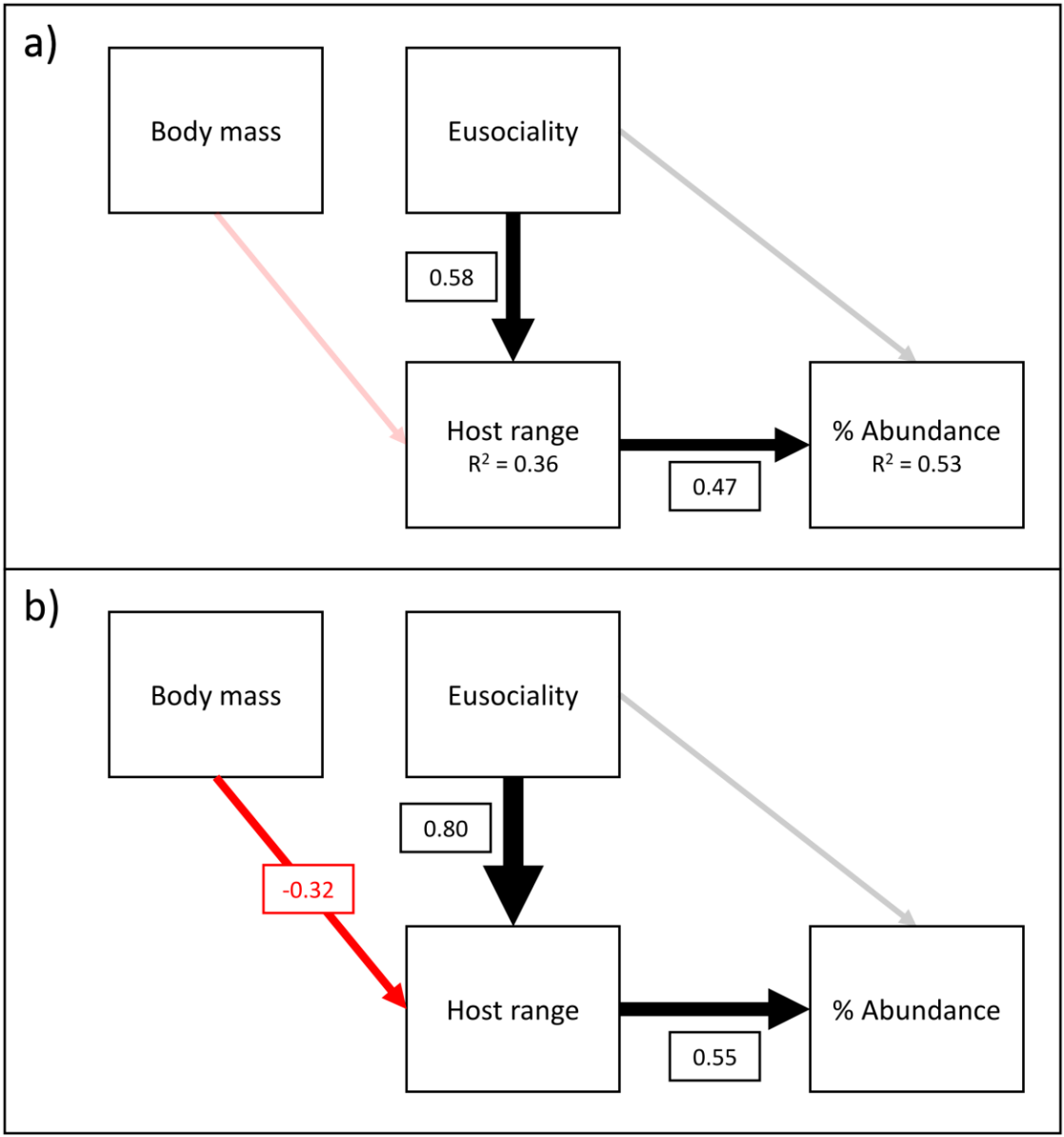

**Figure 2**